\documentclass[english,twocolumn,showpacs,preprintnumbers,amsmath,amssymb,prb,aps]{revtex4}
\usepackage[T1]{fontenc}
\usepackage[latin9]{inputenc}
\usepackage{graphicx}
\usepackage{amssymb}

\makeatletter

\providecommand{\tabularnewline}{\\}

\@ifundefined{textcolor}{}
{%
 \definecolor{BLACK}{gray}{0}
 \definecolor{WHITE}{gray}{1}
 \definecolor{RED}{rgb}{1,0,0}
 \definecolor{GREEN}{rgb}{0,1,0}
 \definecolor{BLUE}{rgb}{0,0,1}
 \definecolor{CYAN}{cmyk}{1,0,0,0}
 \definecolor{MAGENTA}{cmyk}{0,1,0,0}
 \definecolor{YELLOW}{cmyk}{0,0,1,0}
 }

\usepackage{epsfig}

\makeatother

\usepackage{babel}

\begin{document}

\title{Kinetic energy of protons in ice Ih and water: a path integral study}

\author{R. Ram\'{\i}rez$^{a)}$ \let\oldthefootnote\thefootnote\renewcommand{\thefootnote}{{a)}} \footnotetext{Electronic mail: ramirez@icmm.csic.es}\let\thefootnote\oldthefootnote }

\author{C. P. Herrero}

\affiliation{Instituto de Ciencia de Materiales de Madrid (ICMM), Consejo Superior
de Investigaciones Cient\'{\i}ficas (CSIC), Campus de Cantoblanco,
28049 Madrid, Spain }
\begin{abstract}
The kinetic energy of H and O nuclei has been studied by path integral
molecular dynamics simulations of ice Ih and water at ambient pressure.
The simulations were performed by using the q-TIP4P/F model, a point
charge empirical potential that includes molecular flexibility and
anharmonicity in the OH stretch of the water molecule. Ice Ih was
studied in a temperature range between 210-290 K, and water between
230-320 K. Simulations of an isolated water molecule were performed
in the range 210-320 K to estimate the contribution of the intramolecular
vibrational modes to the kinetic energy. Our results for the proton
kinetic energy, $K_{H}$, in water and ice Ih show both agreement
and discrepancies with different published data based on deep inelastic
neutron scattering experiments. Agreement is found for water at the
experimental melting point and in the range 290-300 K. Discrepancies
arise because  data derived from the scattering experiments predict
in water two maxima of $K_{H}$ around 270 K and 277 K, and that $K_{H}$
is lower in ice than in water at 269 K. As a check of the validity
of the employed water potential, we show that our simulations are
consistent with other experimental thermodynamic properties related
to $K_{H}$, as the temperature dependence of the liquid density,
the heat capacity of water and ice at constant pressure, and the isotopic
shift in the melting temperature of ice upon isotopic substitution
of either H or O atoms. Moreover, the temperature dependence of $K_{H}$
predicted by the q-TIP4P/F model for ice Ih is found to be in good
agreement to results of path integral simulations using \emph{ab initio}
density functional theory.
\end{abstract}


\maketitle

\section{Introduction\label{sec:intro}}

The kinetic energy of atomic nuclei in molecules and condensed phases
in thermodynamic equilibrium is related to the interatomic potentials
present in the system. Thus, although in the classical limit the average
kinetic energy of a nucleus does not depend on details of its environment
and interatomic interactions, but only on the temperature (equipartition
principle), in a quantum approach the kinetic energy of a nucleus
gives information on the effective potential in which the particle
moves. 

Deep inelastic neutron scattering (DINS) has been recently applied
to study the thermal average of the kinetic energy of protons, $K_{H}$,
in water and ice. $K_{H}$ is derived from the measured neutron scattering
function by data analysis that includes several numerical corrections,
as the subtraction of the cell signal,\citep{andreani01} the application
of theoretical models as the impulse approximation (that considers
that the neutrons are scattered by single atoms with conservation
of momentum and kinetic energy of the neutron plus the target atom),
and numerical procedures to perform an integral transform of the scattering
function to obtain the momentum distribution of the target atom.\citep{reiter85}
Therefore analysis of experimental DINS data is not free of numerical
limitations that have led in the past to reinterpretation of results.
For example, the effective potential of H along the stretching direction
of the OH bond in supercritical water at 673 K was considered as a
double well potential on the basis of DINS measurements, \citep{reiter04}
however later experiments with longer counting times and subtraction
of the cell signal did not show any evidence for such double well
potential.\citep{pantalei08}

For water, DINS investigations have been performed at atmospheric
pressure and various temperatures: 300 K,\citep{pantalei08} 296 K,\citep{reiter04}
271 K,\citep{pietropaolo08} and 269 K.\citep{pietropaolo08} The
last two temperatures correspond to supercooled water and a substantial
increase in $K_{H}$ of about 58\% (at 271 K) and 40\% (at 269 K)
was reported with respect to ambient water. This large increase was
heuristically interpreted to be a consequence of a double well potential
felt by delocalized H atoms.\citep{pietropaolo08,pietropaolo09} However,
a controversy has arisen about the data analysis at these temperatures,
and it has been claimed that the observed data can be also explained
without invoking large changes in $K_{H}$ and without assuming the
delocalization of H in a double well potential. \citep{soper09} Very
recently a new set of DINS experiments have been performed for water
at temperatures in the range 272-285 K, and seemingly what appears
as a unique set of measurements have been published by two different
groups.\citep{flammini09,pietropaolo09b} From an analysis of their
experimental data these authors found a maximum in $K_{H}$ at about
277 K that was considered as an evidence of a new water structural
anomaly related to the well-known existence of a maximum in the density
of water at the same temperature.\citep{flammini09} DINS derived
values of $K_{H}$ have been also reported at 5 K and at 269 K for
the stable phase of ice at atmospheric pressure 
(ice Ih).\citep{reiter06,reiter04}

The calculation of $K_{H}$ in ice Ih has been done so far with the
help of empirical models consisting of a set of decoupled quantum
harmonic oscillators whose frequency is derived from optical data
and measured vibrational density of states. The contribution of each
harmonic oscillator to $K_{H}$ was determined by the average number
of excited phonons at the given temperature and by an estimation of
the H participation in each normal mode.\citep{moreh10} This numerical
approach predicts that $K_{H}$ in ice Ih is nearly constant between
5 and 300 K. Similar empirical calculations have been performed for
water.\citep{moreh10} A limitation of such approaches is that they
are based on experimental information and neither the pressure nor
the density of the condensed phase can be explicitly considered in
the model. 

Momentum distributions of protons in ice and water have been derived
by path integral (PI) simulations specially designed to sample non-diagonal
density matrix elements. The first attempts used empirical water models
to study ice at 269 K,\citep{burnham06,morrone07} and supercritical
water at 673 K.\citep{morrone07} Water molecules were treated later
by \textit{ab initio} density functional theory (DFT) in studies of
ice at 269 K and water at 300 K,\citep{morrone08} obtaining for the
radial proton momentum distribution closer agreement to experiment.
The thermal average of the proton kinetic energy, $K_{H}$, can be
calculated from the width of its momentum distribution. However, $K_{H}$
is more easily accessible in standard PI simulations, that sample
diagonal elements of the density matrix, by the calculation of the
virial estimator of the kinetic energy.\citep{herman82,parrinello84}
PI simulations of the kinetic energy of atomic nuclei have shown good
agreement to data derived from neutron scattering experiments in solid
neon.\citep{timms96,herrero02}

In the present paper the virial estimator of the kinetic energy, $K,$
has been calculated for water and ice Ih at atmospheric pressure by
PI simulations. The empirical q-TIP4FP/F model has been chosen for
the water simulations because it is an anharmonic flexible potential
that reproduces the experimental density-temperature curve of water
at atmospheric pressure with reasonable accuracy.\citep{habershon09}
This potential model has been recently employed to study the isotopic
shift in the melting temperature of ice Ih.\citep{ramirez10} The
partition of the kinetic energy into H- and O-atom contributions ($K_{H}$,
$K_{O}$) has allowed the comparison to data derived from DINS experiments.
To assess the reliability of our simulations for $K_{H}$ we discuss
also several thermodynamic properties that depend on the kinetic energy,
as the constant pressure heat capacity, $C_{p},$ of ice and water
and isotopic shifts in the melting temperate of ice. An additional
check of the employed potential model has been conducted by comparing
the temperature dependence of $K_{H}$ in ice Ih to results derived
by PI simulations based on \emph{ab initio} DFT calculations using
the SIESTA method.\citep{ordejon96,soler02}

The structure of this paper is as follows. In Sec. \ref{sec:KE} a
short summary of the employed computational conditions is presented.
Then the results obtained with the q-TIP4P/F model for the thermal
average of the kinetic energy of ice, water, and an isolated molecule
are given and compared to available experimental data and also to
\emph{ab initio} DFT simulations. In Sec. \ref{sec:Heat-capacity}
the simulation results of $C_{p}$ at 271 K are presented for ice
and water and compared to experiment. In Sec. \ref{sec:Isotopic-shift}
it is shown that kinetic energy differences between solid and liquid
phases at isothermal conditions determine the sign of the isotopic
shift in the melting temperature. Finally, we summarize our conclusions
in Sec. \ref{sec:conclusions}.

\section{Kinetic energy\label{sec:KE}}

\subsection{Computational conditions\label{sub:Computational-conditions}}

In the PI formulation of statistical mechanics the partition function
is calculated through a discretization of the integral representing
the density matrix. This discretization defines cyclic paths composed
by a finite number $L$ of steps, that in the numerical simulation
translates into the appearance of $L$ replicas (or beads) of each
quantum particle. Then, the implementation of PI simulations relies
on an isomorphism between the quantum system and a classical one,
derived from the former by replacing each quantum particle (here,
atomic nucleus of H and O atoms) by a ring polymer of $L$ classical
particles, connected by harmonic springs with a temperature- and mass-dependent
force constant. Details on this computational method are given elsewhere.\citep{feynman72,gillan88,ceperley95,chakravarty97}
The configuration space of the classical isomorph can be sampled by
a molecular dynamics (MD) algorithm, that has the advantage against
a Monte Carlo method of being more easily parallelizable, an important
fact for efficient use of modern computer architectures. Effective
reversible integrator algorithms to perform PIMD simulations have
been described in detail in Refs. \onlinecite{ma99,tu02,tu98,tuckerman93}.
Ref. \onlinecite{ma96} introduces useful algorithms to treat full
cell fluctuations and multiple time step integration. All calculations
were done using originally developed software and parallelization
was implemented by the MPI library.\citep{pacheco97}

PIMD simulations in the isothermal-isobaric $NPT$ ensemble ($N$
being the number of particles, $P$ the pressure, and $T$ the temperature)
were conducted for ice Ih and water by using the point charge, flexible
q-TIP4P/F model. This model was parameterized to provide the correct
liquid structure, diffusion coefficients, and infrared absorption
frequencies (including the translational and librational regions)
in quantum simulations.\citep{habershon09} Water simulations were
done on cubic cells containing 300 molecules, while ice simulations
included 288 molecules in a proton disordered orthorhombic simulation
cell with parameters $(4a,3\sqrt{3}a,3c)$, with $(a,c)$ being the
standard hexagonal lattice parameters of ice Ih. The total kinetic
energy is defined as $K=2K_{H}+K_{O}$. The kinetic energy of the
H and O nuclei ($K_{H}$, $K_{O}$) were derived by the virial estimator.\citep{herman82,parrinello84,tu98}
We stress that $K_{H}$ and $K_{O}$ are obtained as expected averages
using the exact quantum partition function of the q-TIP4P/F water
model, i.e., our calculation of quantum kinetic energies is free of
any other physical assumption apart from the statistical uncertainty
inherent to the actual PIMD simulations and the use of an empirical
potential model. Expected averages were derived in runs of $5\times10^{5}$
MD steps (MDS) for water and $2.5\times10^{5}$ MDS for ice Ih, using
in both cases a time step of $0.3$ fs. The presence of disorder and
spatial density fluctuations in the liquid causes that larger simulation
runs are required for water. The system equilibration was conducted
in runs of $5\times10^{4}$ MDS. To have a nearly constant precision
in the PI results at different temperatures, the number of beads $L$
was set as the integer number closest to fulfill the relation $LT=6000$
K, i.e., at 210 K the number of beads was $L=28$. Additional computational
conditions are identical to those employed in Ref. \onlinecite {ramirez10}
and they are not repeated here. The simulations were done at atmospheric
pressure in a temperature range of 210-290 K for ice Ih and 230-320
K for water.

Additionally the kinetic energy, $K_{intra}$, associated to the three
vibrational modes of an isolated water molecule was obtained by PIMD
simulations at temperatures between 210-320 K. In the case of an isolated
molecule the virial estimator allows us the calculation of the oxygen,
$K_{O,intra}$, and hydrogen, $K_{H,intra},$ contributions to the
vibrational energy, $K_{intra}.$ The six translational and rotational
degrees of freedom of an isolated water molecule behave classically
at the studied temperatures and their kinetic energy ($k_{B}T/2$
per degree of freedom) has to be summed to $K_{intra}$ to obtain
the total kinetic energy, $K$, of the molecule. If $m_{O}$ and $M$
are the masses of the oxygen atom and the water molecule, then the
fraction of translational kinetic energy corresponding to the oxygen
atom is given by $m_{O}/M$. Analogously for a classical rotation
around a given axis, the fraction of the kinetic energy corresponding
to the O atom is given by $I_{O}/I$, where $I_{O}$ and $I$ are
the moment of inertia of the O atom and the water molecule with respect
to the rotation axis. This ratio has to be calculated for three principal
axes of the molecule to obtain the partition of the rotational kinetic
energy into O and H contributions.\citep{moreh92} The energy unit
used throughout this work (kJ/mol) refers to a mole of either the
molecule (H$_{2}$O) or atom (O or H) under consideration. 

Finally, we have performed PIMD simulations in ice Ih by coupling
our PIMD program to the SIESTA code,\citep{ordejon96,soler02} in
such a way that the potential energy, atomic forces, and stress tensor
used in the $NPT$ simulations are derived from \emph{ab initio} DFT
calculations. The present implementation allows us to study only small
simulation cells, that are not large enough for an appropriate description
of the disorder in the liquid state. We expect to overcome this limitation
in the future by allowing the simultaneous parallelization of both
PI and electronic structure codes. DFT results were derived for a
$2\times1\times1$ supercell of the standard hexagonal cell of ice
Ih with the aim of comparison to the data obtained with the q-TIP4P/F
model. The DFT calculations were done within the generalized-gradient
approximation with the PBE functional.\citep{perdew96} Core electrons
were replaced by norm conserving pseudopotentials\citep{troullier91}
in their fully non-local representation.\citep{kleinman82} In this
study a double-$\zeta$ polarized basis set was used and a grid of
27 k-points was employed for the sampling of the Brillouin zone of
the solid. Integrals beyond two-body interactions are performed in
a discretized real-space grid, its fitness determined by an energy
cutoff of 100 Ry. Average properties were derived by PIMD simulations
runs of $4\times10^{4}$ MDS using a time step of $0.5$ fs. Classical
MD simulations of water using the SIESTA code have been published
in Refs. \onlinecite{marivi96,marivi04,wang11}.

\subsection{Total kinetic energy}

\begin{figure}[!t]
\vspace{-1.0cm}
\hspace*{-0.1cm}
\includegraphics[width= 9cm]{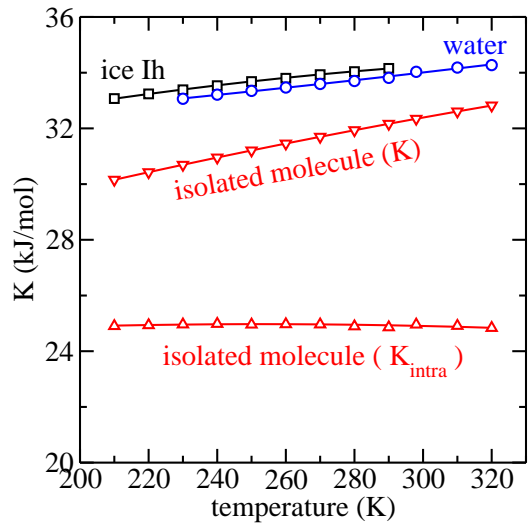}
\vspace{-1.3cm}
\caption{Thermal average of the kinetic energy of
ice Ih and water as a function
of temperature, as calculated by PIMD simulations with the q-TIP4P/F
potential model at $P=1$ atm. The kinetic energy, $K,$ of an isolated
molecule and its vibrational contribution, $K_{intra}$, are also
given. The calculated error bars for ice and water are smaller than
0.01 kJ/mol and for the molecule amount to 0.05 kJ/mol.}
\label{fig:1}
\end{figure}

The temperature dependence of $K$ calculated for ice Ih, water, and
an isolated molecule with the q-TIP4P/F model potential is represented
in Fig. \ref{fig:1}. We observe that at a given temperature $K$
is always larger in ice Ih than in water. For a molecule $K$ is the
sum of global rotational and translational terms (which are classical
contributions at the studied temperatures), and a vibrational term,
$K_{intra}$, corresponding to the two OH stretches and the HOH bending
mode. In Fig. \ref{fig:1} the temperature dependence of the $K_{intra}$
term has been also shown for the isolated molecule. While $K$ in
ice and water increases with temperature, the vibrational term, $K_{intra}$,
of the molecule remains nearly constant. This fact indicates that
the three intramolecular vibrational modes remain mainly in their
ground state in the studied temperature range, while vibrational modes
related to the H-bond network display some amount of thermal population
of excited states. 

The partition of $K$ in ice Ih and water into intra- and intermolecular
contributions ($K_{intra}$ and $K_{inter}$) is of interest to quantify
the effect of the presence of H-bonds in this observable. The intramolecular
term, $K_{intra}$, is attributed to the internal vibrational modes
of the molecule, while the intermolecular one, $K_{inter}$, is related
to the librational (rotational) and translational motions of the water
molecules in the condensed phases. The presence of a H-bond network
causes a large broadening of the internal vibrational modes of the
water molecule and a shift in the frequencies of their peaks in the
infrared absorption spectra.\citep{habershon09} This fact implies
that both inter- and intramolecular motions are clearly coupled in
the condensed phases of water. Therefore the partition of the total
kinetic energy is presented here only as an approximation for didactic
purposes in ice and water, and it is not intended to suggest that
both types of degrees of freedom may be treated as separable variables. 

We estimate $K_{intra}$ in ice Ih and water with the help of a proportionality
relation based on the fact that at the studied temperatures $K_{intra}$
is close to the kinetic energy of the ground state associated to three
intramolecular modes,$\; K_{GS,intra}$. Particularizing to the case
of ice we have,

\begin{equation}
\frac{K_{intra}(ice)}{K_{intra}(molecule)}\approx\frac{K_{GS,intra}(ice)}{K_{GS,intra}(molecule)}\;.\label{eq:intra}\end{equation}
As a rough estimation of the ground state kinetic energy, $K_{GS,intra}$,
in either ice, water or an isolated molecule, we employ here a quasi-harmonic
approximation, so that a vibrational mode of wavenumber $\omega$
displays a kinetic energy of $\hslash\omega/4$ in its ground state.
Thus $K_{GS,intra}$ will be approximated as 

\begin{equation}
K_{GS,intra}\approx\frac{\hslash}{4}\left(2\omega_{OH}^{QA}+\omega_{HOH}\right)\;.\label{eq:intra2}\end{equation}
The wavenumber $\omega_{OH}^{QA}$ is used here as an estimation of
the average wavenumber of the two stretching modes (symmetric and
asymmetric) of an isolated water molecule or the corresponding vibrational
bands in the condensed phases. $\omega_{OH}^{QA}$ is determined from
the equilibrium bond distance, $d_{OH}$, of either ice, water or
an isolated molecule, by calculating the second derivative of the
potential energy with respect to $d_{OH}$, and by considering the
actual O and H masses.\citep{ramirez10} The bending wavenumber is
$\omega_{HOH}=$1600 cm$^{-1}$. We recall that the bending potential
is described by a simple harmonic term in the employed water model.\citep{habershon09}
The ground state assumption for $K_{intra}$ in Eq. (\ref{eq:intra})
is further justified by noting that the excitation energy corresponding
to the vibrational mode of lowest frequency, $\hslash\omega_{HOH},$
is about seven times larger than the thermal energy, $k_{B}T$, at
the highest studied temperature (320 K), and thus the thermal population
of excited vibrational modes, as given by a Bose-Einstein statistics,\citep{moreh92,andreani01b}
results vanishingly small.

\begin{table*}
\caption{Thermal average of the kinetic energy, $K,$ of water and ice
Ih at
atmospheric pressure and of an isolated water molecule. The results
were derived by PIMD simulations at 270 K. The energy $K$ has been
partitioned into O- and H-atom contributions ($K_{O},$ $K_{H}$).
For water and ice the estimated intramolecular and intermolecular
contributions to $K$, $K_{O}$, and $K_{H}$ are also given. For
a molecule the classical translational and rotational energy is given
as $K_{tr+rot}$. This classical term has been partitioned into O
and H components. $d_{OH}$ is the thermal average of the intramolecular
OH distance, and $\omega_{OH}^{QA}$ is the quasi-harmonic stretching
wavenumber calculated at the distance $d_{OH}$. The last column
displays
the theoretical results of Ref. \onlinecite{moreh10} for ice Ih at
269 K. Kinetic energies are given in kJ/mol (to convert to meV multiply
by a factor of 10.37). The standard error in the final digit is given
in parenthesis.}

\centering{}\label{tab:1}\begin{tabular}{lcccc}
\hline
 & molecule & water & ice Ih & ice Ih%
\footnote{Reference \onlinecite{moreh10}.%
}\tabularnewline
\hline
$K$ & 31.69(5) & $\quad$33.59(1)$\quad$ & 33.93(1) &
33.38\tabularnewline
$K_{intra}$  & 24.96(5) (79\%) & 23.5 (70\%) & 23.2 (68\%) &
-\tabularnewline
$K_{inter}$  & -- & 10.1 (30\%) & 10.7 (32\%) & -\tabularnewline
$K_{tr+rot}$ & 6.73 (21\%) & -- & -- & \tabularnewline
\hline
$K_{O}$ & 4.68(1) & 5.04(1) & 5.09(1) & 5.01\tabularnewline
$K_{O,intra}$  & 1.53(1) (33\%) & 1.4 (29\%) & 1.4 (27\%) &
-\tabularnewline
$K_{O,inter}$  & -- & 3.6 (71\%) & 3.7 (73\%) & -\tabularnewline
$K_{O,tr+rot}$ & 3.16 (67\%) & -- & -- & \tabularnewline
\hline
$K_{H}$ & 13.50(2) & 14.28(1) & 14.42(1) & 14.18\tabularnewline
$K_{H,intra}$  & 11.72(2) (87\%) & 11.0 (77\%) & 10.9 (76\%) & 11.2
(79\%)\tabularnewline
$K_{H,inter}$  & -- & 3.3 (23\%) & 3.5 (24\%) & 3.0
(21\%)\tabularnewline
$K_{H,tr+rot}$ & 1.78 (13\%) & -- & -- & \tabularnewline
\hline
$d_{OH}$ (\AA) & 0.957 & 0.978 & 0.982 & \tabularnewline
$\omega_{OH}^{QA}$ (cm$^{-1}$) & 3688 & 3424 & 3375 & \tabularnewline
\hline
\end{tabular}
\end{table*}

The partition of $K$ into $K_{intra}$ and $K_{inter}$ contributions
is presented in Table \ref{tab:1} at a temperature of 270 K. For
water the intermolecular contribution, $K_{inter}$, is estimated
to be 30\% (10.1 kJ/mol) of $K$, while for ice Ih this contribution
is slightly larger (32\%, 10.7 kJ/mol), in agreement with the expectation
that H-bonds in ice are stronger than in water.\citep{chapling10}
We display also the average OH bond distance, $d_{OH}$, and the corresponding
quasi-harmonic stretching wavenumber, $\omega_{OH}^{QA}$. The bond
distance increases in the order molecule $\ll$ water < ice. We note
that x-ray Compton scattering studies of water and ice Ih have shown
that the elongation of the OH bond in ice Ih is correlated to its
stronger H-bond network,\citep{nygard06} in agreement to the simulation
results for $d_{OH}$ and $K_{inter}$ shown in Tab. \ref{tab:1}.
Our result of $K$ in ice Ih displays a reasonable agreement to a
previous calculation that is based on the use of experimental vibrational
frequencies assuming a harmonic approximation (see the last column
of Tab. \ref{tab:1}).\citep{moreh10}

\subsection{Kinetic energy of O and H nuclei }

\begin{figure}
\vspace{-1.8cm}
\hspace*{0.8cm}
\includegraphics[width= 9cm]{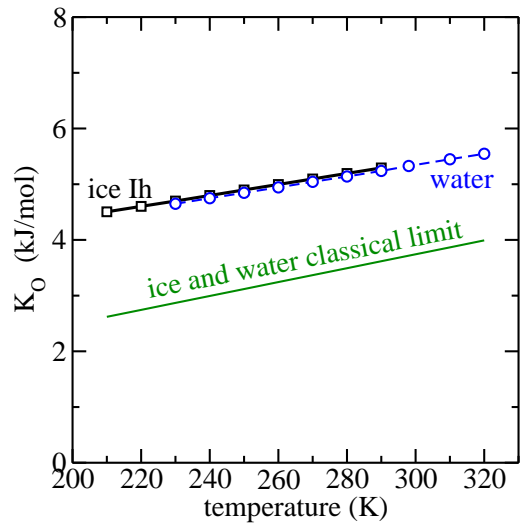}
\vspace{-0.5cm}
\caption{Oxygen contribution, $K_{O}$, to the
kinetic energy of ice Ih and
water as a function of temperature. The classical limit for ice and
water amounts to (3/2)$k_{B}T$ per atom. The calculated error bars
are smaller than 3$\times10^{-3}$ kJ/mol. For a given temperature,
the difference of $K_{O}$ in ice Ih and water ($\sim0.05$ kJ/mol)
is significantly larger than the error bar.}
\label{fig:2}
\end{figure}

The O nuclei contribution, $K_{O}$, to the kinetic energy is presented
in Fig. \ref{fig:2}. $K_{O}$ in the condensed phases is significantly
larger than its classical limit, i.e., quantum effects related to
the oxygen mass are relevant for its kinetic energy. At a given temperature
$K_{O}$ is slightly larger in ice Ih than in water (see also the
$K_{O}$ values of ice and water at 270 K in Table \ref{tab:1}).

\begin{figure}
\vspace{-1.8cm}
\hspace*{0.7cm}
\includegraphics[width= 9cm]{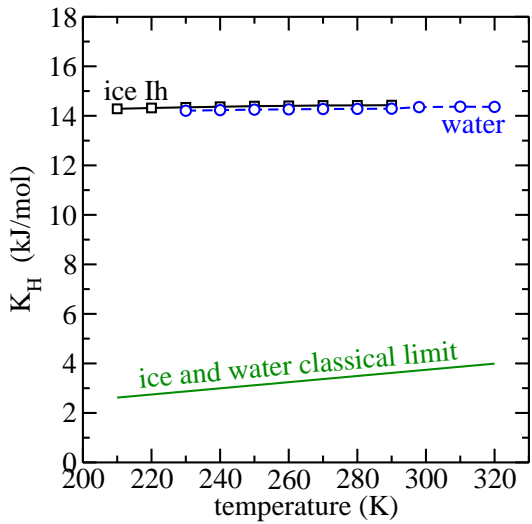}
\vspace{-0.5cm}
\caption{Proton contribution, $K_{H}$, to the
kinetic energy of ice Ih and
water as a function of temperature. The classical limit for the condensed
phases corresponds to (3/2)$k_{B}T$ per atom. The calculated error
bars are smaller than $5\times10^{-3}$ kJ/mol. At constant temperature,
the difference of $K_{H}$ in ice Ih and water ($\sim0.1$ kJ/mol)
is significantly larger than the error bar.}
\label{fig:3}
\end{figure}

The temperature dependence of $K_{H}$ is presented in Fig. \ref{fig:3}
for ice and water. The $K_{H}$ contribution to the kinetic energy
is substantially larger than the $K_{O}$ term and it is also much
larger than the classical limit, i.e., quantum effects related to
their low atomic mass are very important for H atoms. At a given temperature
$K_{H}$ is always larger in ice Ih than in water.

The partition of $K_{H}$ and $K_{O}$ into intramolecular and intermolecular
contributions can be derived by assuming the following proportionality
relation for the intramolecular kinetic energy in the condensed phases

\begin{equation}
\frac{K_{H,intra}}{K_{O,intra}}\approx\left(\frac{K_{H,intra}}{K_{O,intra}}\right)_{molecule}\;,\label{eq:intra_ratio}\end{equation}
 i.e., the ratio $K_{H,intra}/K_{O,intra}$ at a given temperature
is assumed to be constant $(\sim7.7)$ for ice, water, and an isolated
molecule. Although we can not provide a theoretical estimation of
the error of this assumption, water molecules remain as clearly recognizable
entities in the condensed phases at atmospheric pressure, and therefore
the intermolecular interactions are not expected to change drastically
the constant ratio assumed by Eq. (\ref{eq:intra_ratio}). In Table
\ref{tab:1} the result of this partition is given at 270 K. For water,
the intermolecular contribution to $K_{O}$ is larger than the intramolecular
term (71\% and 29\%, respectively). However in the case of $K_{H}$
in water, the intermolecular contribution (23\%, 3.3 kJ/mol) is much
lower than the intramolecular term (77\%, 11 kJ/mol, corresponding
to the two OH stretches and the HOH bending mode). For ice Ih, having
a stronger H-bond network than water,\citep{chapling10} we find that
intermolecular contributions to $K_{O}$ and $K_{H}$ are slightly
larger than in water, while the intramolecular contributions are slightly
lower.

\subsection{Comparison to DINS data}

\begin{figure}
\vspace{-1.0cm}
\hspace*{-0.1cm}
\includegraphics[width= 9cm]{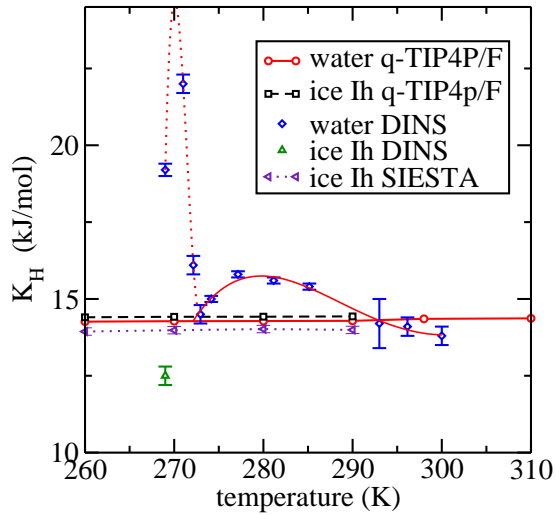}
\vspace{-1.3cm}
\caption{PIMD results obtained with the q-TIP4P/F
model for the proton kinetic
energy in water and ice Ih are compared to data derived from DINS
experiments at temperatures in the range 269-300
K.\citep{reiter04,pantalei08,reiter06,pietropaolo08,flammini09,pietropaolo09b}
Also PIMD results for ice Ih derived by \emph{ab initio} DFT calculations
with the SIESTA code are given. Lines are guides to the eyes.}
\label{fig:4}
\end{figure}

In Fig. \ref{fig:4} our $K_{H}$ results are compared to data derived
from DINS experiments.\citep{reiter04,pantalei08,reiter06,pietropaolo08,flammini09,pietropaolo09b}
Our calculation for water agrees reasonably well with the data derived
from the neutron measurements at 273 K and the three values in the
range 290-300 K. However, significant deviations are found in the
temperature ranges between 269-272 K and 274-286 K. 

The value of $K_{H}$ derived at 271 K from the DINS experiment predicts
an excess of 58\% (8 kJ/mol) with respect to the room temperature
result. This excess was associated to a coherent delocalization of
the proton between neighbor oxygen atoms.\citep{pietropaolo09b,flammini09}
However, this tentative explanation has been criticized by noting
that measured proton momentum distribution implies a contracted proton
wave function which is inconsistent with the purported increased proton
delocalization. \citep{soper09} Moreover, coherent delocalization
(i.e., tunneling) of a proton is associated to a resonance of the
particle between two localized states. A coherent delocalization of
the proton at $T_{c}=$271 K would imply that the resonance frequency
associated to the coherent process must be large enough to display
the following relation to the thermal energy: $\hslash\omega_{c}\gg k_{B}T_{c}$.
In other case the quantum coherence would be destroyed by thermal
excitations. This inequality should remain valid if $T_{c}$ increases
to 273 K, i.e., a slight raise of 2 K should not provide enough thermal
energy to destroy a coherent process of wavenumber $\omega_{c}$ for
the proton. Therefore, it seems somewhat contradictory that the excess
of $K_{H}$ attributed to a coherent process at 271 K disappears completely
at 273 K. 

An alternative explanation for the large value of $K_{H}$ as a localized
proton state at 271 K leads also to an unphysical picture. In this
respect, as long as the water molecules maintain their molecular structure,
i.e., without a dramatic increase in their stretching frequencies
(which has been never observed), the intramolecular contribution to
$K_{H}$ must remain below the isolated molecule result of about 11
kJ/mol (see Fig. \ref{fig:3}). Therefore an excess as large as 8
kJ/mol in the intramolecular energy seems to be incompatible with
the experimental molecular structure of water. 

\begin{figure}
\vspace{-1.0cm}
\hspace*{-0.1cm}
\includegraphics[width= 9cm]{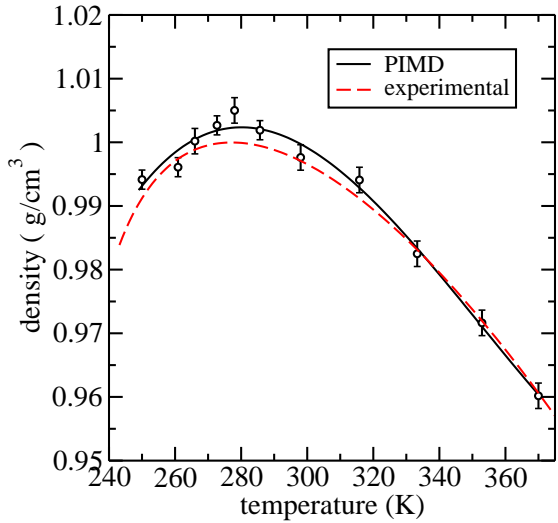}
\vspace{-1.3cm}
\caption{Density of water at 1 atm obtained from
PIMD simulations and experimental
data from Ref. \onlinecite{kell75}. Simulation results correspond
to runs of $4\times10^{6}$ MDS. The continuous line is a cubic fit
to the simulation data.}
\label{fig:5}
\end{figure}

The second maxima in $K_{H}$ derived from the DINS experiments in
the temperature range 274-286 K has been attributed to a new structural
anomaly of water related to the existence of a maximum in the water
density as a function of temperature.\citep{flammini09,pietropaolo09b}
This explanation is however not supported by our simulations. The
employed water potential is able to reproduce realistically the experimental
density maximum of water as a function of temperature 
(see Fig. \ref{fig:5}).\citep{habershon09}
In the temperature range between 274-286 K there appears a maximum
in the density of water. Both experimental and simulation data agree
in the fact that the water density in this temperature range changes
by less than 1/1000. Our simulations predict that such a small density
change does not affect $K_{H}$ in any significant way (see Fig. \ref{fig:4}).
However, the excess in $K_{H}$ at 277 K as derived from the DINS
experiments amounts to 2.2 kJ/mol with respect to the room temperature
value. This excess energy is rather large, and of similar magnitude
to the total H-bond network contribution to $K_{H},$ that amounts
to about 3.3 kJ/mol for the employed potential model (see Table \ref{tab:1}
for our result of $K_{H,inter}$ in water at 270 K, this value remains
nearly constant in the studied temperature range). Thus an excess
of 2.2 kJ/mol represents, for the employed water model, an increase
of 66\% in the kinetic energy associated to the intermolecular interaction.
We believe that such a large energetic modification should imply a
dramatic change of the H-bond network that however has not been observed
in any other spectroscopic or thermodynamical data of water around
277 K.

Another discrepancy between our simulations and the DINS analysis
of $K_{H}$ is that we find that at isothermal conditions $K_{H}$
is about 0.14 kJ/mol larger in ice than in water. However the DINS
value reported for ice\citep{reiter06} at 269 K is much lower ($\sim$
7 kJ/mol) than that of water\citep{pietropaolo08} at the same temperature. 

It is difficult to find a definite explanation for the agreement found
between simulation data for $K_{H}$ in water and those derived from
neutron scattering at some temperatures (273 K and 290-300 K) and
the discrepancies found at other temperatures (269-272 K and 274-286
K) and in the result for ice Ih. Both theoretical and experimental
approaches include basic assumptions that might lead to unphysical
results. From a theoretical side, the main limitation comes from the
use of an empirical water potential. In particular, any dynamic process
involving delocalization of protons between different molecules (i.e.,
breaking and forming of intramolecular OH bonds) can not be described
by the employed q-TIP4P/F potential. From the experimental side, the
bibliography shows an experimental controversy concerning the validity
of the models employed for the data analysis of the DINS experiments
at 269 and 271 K.\citep{soper09,pietropaolo09} Moreover, the numerical
data analysis of the diffraction experiment implies corrections due
to the experimental setup (detector position, subtraction of the cell
signal) and a numerical integral transform that might be subject to
numerical imprecisions. It has been mentioned in the Introduction
that numerical reinterpretation of DINS experiments have already happened
in recent times.\citep{reiter04,pantalei08} 

In the face of this disagreement, the rest of the paper is intended
as a consistency check for the ability of the employed empirical potential
to reproduce the kinetic energy of the atoms in ice Ih and water.
Firstly, we go beyond the empirical potential by presenting results
of \emph{ab initio} DFT simulations.

\subsection{Beyond the empirical potential}

In Fig. \ref{fig:4} we show the $K_{H}$ results for ice Ih derived
from PIMD simulations using atomic forces calculated with the SIESTA
code.\citep{ordejon96,soler02} The temperature dependence of $K_{H}$
predicted by the q-TIP4P/F model shows good agreement to the \emph{ab
initio} DFT results. The main difference between both sets of results
is that values predicted by the q-TIP4P/F model are $\sim3\%$ larger
than the DFT ones. The origin of this shift is related to a different
description of the OH stretching, as evidenced by a normal mode analysis
that provides harmonic stretching frequencies about $4\%$ larger
in the case of the q-TIP4P/F model. The \emph{ab initio} data were
derived in a small simulation cell (see Sec. \ref{sub:Computational-conditions}).
Thus, the finite size error has been estimated by comparing $K_{H}$
values obtained in a small $(2a,a,c)$ and a large $(4a,3\sqrt{3}a,3c)$
supercell (i.e., containing 8 and 288 water molecules, respectively)
by using the q-TIP4P/F model. This finite size effect is rather low,
the smaller cell shows a $K_{H}$ value $0.3\%$ lower than the larger
one. We stress that the q-TIP4P/F potential was parameterized to give
correct structure, diffusion coefficient, and infrared absorption
frequencies in the \emph{liquid} \emph{state.} \citep{habershon09}
Therefore the reasonable agreement to the \emph{ab initio} data found
for $K_{H}$ in ice Ih should be considered as a realistic prediction
of the empirical potential for a property that was not implicitly
built in by the potential parameterization.

In the following Sections we continue our check of the q-TIP4P/F potential
by comparing some thermodynamical properties (heat capacity and isotopic
shifts) that depend on the kinetic energy of the nuclei with experimental
data.

\section{Heat capacity at 271 K\label{sec:Heat-capacity}}

\begin{table}
\vspace*{-0.2cm}
\caption{Experimental results for $C_{p}$ of water (Ref.
\onlinecite{angell82})
and ice Ih (Ref. \onlinecite{feistel05}) are compared with PIMD results
at 271 K and 1 atm. The calculated $C_{p}$ has been decomposed into
potential ($C_{p1}$ and $C_{p2}$) and kinetic ($C_{p3}$ and $C_{p4}$)
energy contributions. The numerical estimation of $C_{p4}$ from the
data derived from DINS
experiments\citep{pietropaolo08,flammini09,pietropaolo09b}
is also given. The employed unit is J~mol$^{-1}$K$^{-1}$.}
\label{tab:2}\begin{tabular}{lcc}
\hline
 & water & ice\tabularnewline
\hline
$C_{p}$ experimental & 76.1 & 37.4 \tabularnewline
$C_{p}$ PIMD & 71.4$\pm2$ & 36.1$\pm0.8$\tabularnewline
$C_{p1}(V_{intra})$ & -9.0$\pm0.4$ & -7.4$\pm0.2$\tabularnewline
$C_{p2}(V_{inter})$ & 66.3$\pm1$ & 31.6$\pm0.5$\tabularnewline
$C_{p3}(K_{O})$ & 10.3$\pm0.2$$\quad$ & 9.9$\pm0.05$\tabularnewline
$C_{p4}(K_{H})$ & 3.7$\pm0.2$ & 2.0$\pm0.05$\tabularnewline
$C_{p4}(K_{H},$ DINS) & $-10^{4}$ & -\tabularnewline
\hline
\end{tabular}
\end{table}

The enthalpy, $H$, calculated with the employed water model can be
expressed as\begin{equation}
H=V_{intra}+V_{inter}+K_{O}+K_{H}+PV\;,\label{eq:H}\end{equation}
where the first four summands give the internal energy as a sum of
potential ($V$) and kinetic energy ($K$) terms and the potential
energy is partitioned into a sum of intramolecular ($V_{intra})$
and intermolecular ($V_{inter})$ contributions. We have fitted our
PIMD results for the enthalpy of ice and water to a quadratic function
of temperature in the range 250-290 K. Then the constant pressure
heat capacity

\begin{equation}
C_{p}=\left(\frac{\partial H}{\partial T}\right)_{P}\;\end{equation}
has been calculated at 271 K by performing the temperature derivative
of the fitted function. The results are presented in Table \ref{tab:2}
and compared to experimental data. The employed water model provides
realistic values of $C_{p}$ for both water and ice at 271 K. The
deviation with respect to experiment amounts to about 6\% in water
and to 4\% in ice Ih, with simulated values lower than experimental
ones. Note that the experimental heat capacity of ice Ih is nearly
one half of that of water, a fact that is reasonably reproduced by
the simulations. This overall agreement encourages us to analyze separately
the potential ($C_{p1},C_{p2}$) and kinetic ($C_{p3},C_{p4}$) energy
contributions to $C_{p}$, as derived from the temperature derivative
of the first four summands in Eq. (\ref{eq:H}). At atmospheric pressure
the temperature derivative of the $PV$ term gives a vanishingly small
contribution to $C_{p}$. 

The four contributions to the calculated $C_{p}$ have been summarized
in Table \ref{tab:2}. We note that the contribution $C_{p1}$ of
the intramolecular potential energy ($V_{intra}$) is negative for
both water and ice Ih. This result is an anharmonic effect related
to the reduction of the $d_{OH}$ distance with increasing temperature,
a fact that has been experimentally reported in water,\citep{eck58,nygard06}
and that is reproduced by the employed model potential. The largest
contribution ($C_{p2}$) to the heat capacity comes from the potential
energy of the H-bond network, $V_{inter}$. $C_{p2}$ amounts to 93\%
of the calculated $C_{p}$ in water and to 88\% in ice, a result in
line to the expectation that the high heat capacity of water is a
consequence of its H-bond network. Kinetic energy contributions ($C_{p3},C_{p4}$)
are comparatively much lower. $C_{p3}$, derived from the temperature
derivative of the kinetic energy $K_{O}$, is several times larger
than $C_{p4}$, that is derived from $K_{H}$. In particular, we find
that $C_{p4}$ amounts to less than 4 J mol$^{-1}$K$^{-1}$ (5\%
of the calculated $C_{p}$) in water. 

The last line of Table \ref{tab:2} summarizes the value of $C_{p4}$,
as derived from the temperature derivative of the $K_{H}$ values
obtained from the DINS experiments, that were plotted in Fig. \ref{fig:4}.
This contribution is negative and its absolute value ($\sim10^{4}$J
mol$^{-1}$K$^{-1}$) is several orders of magnitude larger than the
experimental heat capacity. Both sign and magnitude of this estimation
of $C_{p4}$ seem to be unphysical and incompatible with the experimental
$C_{p}$ data at 271 K. This fact let us to question if the maximum
in $K_{H}$ derived from the DINS analysis around 270 K (see Fig.
\ref{fig:4}) corresponds to any real physical effect.

\section{Isotopic shift in the melting temperature \label{sec:Isotopic-shift}}

\begin{figure}
\vspace*{-2.9cm}
\hspace*{-0.1cm}
\includegraphics[width= 9cm]{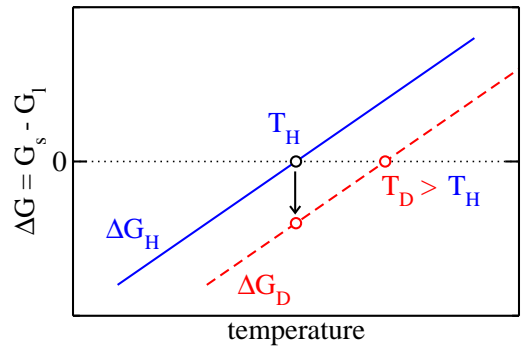}
\vspace{-1.9cm}
\caption{Schematic representation of the free energy difference,
$\Delta G$,
between solid and liquid phases of water as a function of temperature
at constant pressure. The continuous line represents $\Delta G_{H}$
for phases containing the isotope $^{1}$H and the melting temperature
of the solid is $T_{H}$. Upon isotopic substitution of H by deuterium,
the free energy curve is shifted to the broken one, $\Delta G_{D}$.
At given temperature the shift of $\Delta G_{D}$ with respect to
$\Delta G_{H}$ (shown by an arrow at the temperature $T_{H}$) can
be calculated by the definite integral in Eq. (\ref{eq:delta_G}).
A normal isotopic shift implies that the melting temperature of the
heavier isotope is higher ($T_{D}>T_{H}$).}
\label{fig:6}
\end{figure}

For ice Ih at atmospheric pressure experimental isotopic shifts indicate
that heavier isotopes display higher melting temperatures. Thus deuterated
ice Ih melts at a temperature 3.8 K higher than normal ice, while
the isotopic shift for tritiated ice amounts to 4.5 K. Also H$_{2}$$^{18}$O
melts at a temperature 0.3 K higher than ice Ih made of normal H$_{2}$$^{16}$O.
Isotopic shifts in the melting temperature are a consequence of the
dependence of the Gibbs free energy, $G$, on the isotope mass. At
constant ($T$,$P)$, this dependence results to be a function of
the kinetic energy of the isotope. Thus if $G_{H}$ is the Gibbs free
energy of a water phase containing $^{1}$H atoms, the free energy
of the corresponding deuterated phase, $G_{D}$, is given by the following
relation\citep{ramirez10}

\begin{equation}
G_{D}=G_{H}-\int_{m_{H}}^{m_{D}}\frac{dm}{m}K_{H}(m)\;,\label{eq:GD}\end{equation}
where $m_{H}$ and $m_{D}$ are the H and deuterium (D) masses, and
$K_{H}(m)$ is the thermal average of the kinetic energy of a H isotope
with mass $m$. Note that the mass $m$ appears as an auxiliary variable
in Eq. (\ref{eq:GD}) and corresponds to a real isotope only at the
integration limits ($m_{H},m_{D}$). The Gibbs free energy difference,
$\Delta G=G_{s}-G_{l}$, between the solid ($s$) and liquid ($l)$
phases at constant $(T,P)$ can be expressed from Eq. (\ref{eq:GD})
as

\begin{equation}
\Delta G_{D}=\Delta G_{H}-\int_{m_{H}}^{m_{D}}\frac{dm}{m}\Delta K_{H}(m)\;,\label{eq:delta_G}\end{equation}
where

\begin{equation}
\Delta K_{H}(m)=K_{H,s}(m)-K_{H,l}(m)\;.\end{equation}
$K_{H,s}(m)$ is the isotope kinetic energy in the solid and $K_{H.l}(m)$
the corresponding quantity in the liquid phase. A schematic representation
of the free energy difference $\Delta G_{H}$ at constant pressure
is shown in Fig. \ref{fig:6} as a function of $T$. At the melting
temperature, $T_{H}$, the free energy of both phases must be identical
and $\Delta G_{H}$ vanishes. At temperatures below $T_{H}$ we find
that $\Delta G_{H}<0$, i.e., the solid is the thermodynamically stable
phase as it displays lower free energy than the liquid. However, for
$T>T_{H}$ one finds $\Delta G_{H}>0$ and then the stable phase is
the liquid one. The isotope effect in the melting temperature can
be obtained by calculating $\Delta G_{D}$ with the help of Eq. (\ref{eq:delta_G}).
A normal isotope effect implies that the definite integral in Eq.
(\ref{eq:delta_G}) takes a positive value. Thus at constant temperature
$\Delta G_{D}$ becomes smaller than $\Delta G_{H}$ obtaining the
schematic result represented in Fig. \ref{fig:6}. Note that the melting
point for the heavier isotope, $T_{D}$ (where $\Delta G_{D}\equiv0$),
is shifted towards higher temperatures. 

We conclude that it is the sign associated to the value of the definite
integral in Eq. (\ref{eq:delta_G}) what determines if the isotope
effect in the melting temperature is normal (i.e., higher melting
temperature for a heavier isotope mass if the sign is positive) or
it is an inverse isotope effect (lower melting point for a heavier
isotope mass if the sign is negative). A sufficient (but not necessary)
condition to obtain a normal isotope effect is that 

\begin{equation}
\Delta K_{H}(m)>0;\quad\mathrm{for\;}m_{H}\leq m\leq m_{D}\label{eq:K}\end{equation}
i.e., the isotope kinetic energy is larger in the solid than in the
liquid phase for the isotope masses $m$ in the studied interval.

Our simulations show that the condition in Eq. (\ref{eq:K}) is satisfied
for the isotopic substitution of H by either deuterium or tritium.
The analogous condition, $\Delta K_{O}(m)>0$, is also satisfied for
the isotopic substitution of $^{16}$O by $^{18}$O, with $m(^{16}\mathrm{O})\leq m\leq m(^{18}\mathrm{O})$.
Thus the employed water model predicts normal isotopic shifts in the
melting temperature upon isotopic substitution of either $^{1}$H
or $^{16}$O by a heavier isotope, in agreement to experiment. Detailed
results for deuterated and tritiated phases have been presented elsewhere.\citep{ramirez10}
The values collected in Table \ref{tab:1} at 270 K show that for
hydrogen we get the value $\Delta K_{H}(m_{H})=0.14$ kJ/mol, i.e.,
$K_{H}$ is larger in the solid than in the liquid, while for $^{16}$O
we find than $K_{O}$ is 0.05 kJ/mol larger in ice than in water. 

The $K_{H}$ results derived from DINS data for water and ice Ih at
269 K give $\Delta K_{H}(m_{H})=-6.7$ kJ/mol, i.e., $K_{H}$ is much
lower in ice than in water. Note that here the inequality of Eq. (\ref{eq:K})
is not satisfied for $m_{H}$. This fact implies that a small increase
in the mass of hydrogen would produce an inverse isotope effect in
the melting temperature. This behavior for $m_{H}$ does not imply
an inverse isotope effect for deuterium, because the latter effect
is determined by the definite integral in Eq. (\ref{eq:delta_G})
that involves all masses $m$ with $m_{H}\leq m\leq m_{D}$. However,
the value of $\Delta K_{H}(m_{H})=-6.7$ kJ/mol at 269 K derived from
the DINS analysis contrasts in sign and magnitude with our result
of $\Delta K_{H}(m_{H})=0.14$ kJ/mol. We stress that the employed
potential model provides a realistic prediction of the isotopic shift
in the melting temperature of ice, a fact that let us expect that
our calculation of $\Delta K_{H}(m_{H})$ must be also reasonably
realistic.\citep{ramirez10} This consideration rises further doubts
about the physical significance of the maximum in $K_{H}$ derived
from the DINS analysis for liquid water around 270 K.

\section{Conclusions\label{sec:conclusions}}

Quantum PIMD simulations at atmospheric pressure show that the q-TIP4P/F
model provides realistic results for several thermodynamic properties
of water and ice Ih. In particular, satisfactory agreement to experiment
has been found in the dependence of the water density with temperature,\citep{habershon09}
the isotopic shifts in the melting temperature of deuterated and tritiated
ice,\citep{ramirez10} and the constant-pressure heat capacity of
water and ice Ih at 271 K. In the present paper we have focused on
the calculation of the proton kinetic energy, $K_{H}$, and its comparison
to available data derived from the analysis of neutron scattering
experiments,\citep{reiter04,pantalei08,reiter06,pietropaolo08,flammini09,pietropaolo09b}
and also to results based on \emph{ab initio} DFT simulations.

This comparison has shown a remarkable agreement between the computational
model and DINS data for water at some temperatures (273 K and in the
range 290-300 K), but a strong disagreement at other similar temperatures
(in the range 269-272 K and 274-286 K). Our simulations predict that
$K_{H}$ remains nearly constant in the temperature range where DINS
data are available at atmospheric pressure. On the contrary, $K_{H}$
values derived from the scattering experiments display a remarkable
temperature dependence with two maxima at 270 K and 277 K. This strong
temperature dependence and, particularly, the large DINS value reported
for $K_{H}$ at 271 K (58\% larger than at room temperature) are the
main discrepancies found with our results for water. Our simulations
predict that under isothermal conditions $K_{H}$ is larger in ice
than in water while the values derived from DINS data at 269 K forecast
that $K_{H}$ is much lower in ice Ih than in water. 

In the face of the discrepancies found between our simulation results
for $K_{H}$ and those derived from neutron scattering experiments,
we have related $K_{H}$ to other thermodynamic properties pursuing
to gain a deeper understanding of the implication of such differences.
We find that the maximum in the density of water is not related to
any significant effect in the proton kinetic energy at atmospheric
pressure. In the analysis of $C_{p}$ our simulations display reasonable
results for both water and ice Ih at 271 K. The $K_{H}$ contribution
to $C_{p}$ amounts to about 4 J mol$^{-1}$K$^{-1}$ in water (5\%
of $C_{p}$ ). However, the temperature dependence of $K_{H}$ as
derived from DINS experiments leads to a dramatic negative contribution
to $C_{p}$ of about $-10^{4}$ J mol$^{-1}$K$^{-1}$. This value
can not be directly compared to experimental data, but both its sign
and magnitude seem to be outside any reasonable physical boundary.
At this point it is clear that further independent experiments and
theoretical calculation are necessary to clarify this question.

The calculated values of $K_{O}$ and $K_{H}$ in water and ice at
isothermal conditions show that the largest results are always associated
to the solid phase. This fact determines the sign of the isotopic
shifts in the melting temperature upon isotopic substitution of either
H or O atoms. In both cases the model predictions are in agreement
to experiment. The temperature dependence of $K_{H}$ in ice Ih has
shown satisfactory agreement to results derived from PIMD simulations
based on\emph{ ab initio }DFT calculations using the SIESTA code.\citep{ordejon96,soler02} 

Summarizing, it is clear that an empirical potential model of water
can not reproduce quantitatively all thermodynamic properties of the
real system. However, it is the combined analysis of several thermodynamic
properties such as the proton kinetic energy $K_{H}$, the intramolecular
and intermolecular contributions to $K_{H}$, the heat capacity $C_{p}$,
the isotopic shifts in melting temperatures, and the density-temperature
curve of water, what let us believe that the employed model provides
realistic results for the temperature dependence of $K_{H}$ in water
and ice Ih at atmospheric pressure. 

\acknowledgments 

This work was supported by Ministerio de Ciencia e Innovación (Spain)
through Grant No. FIS2009-12721-C04-04 and by Comunidad Autónoma de
Madrid through project MODELICO-CM/S2009ESP-1691. We thank E. Anglada
and J.M. Soler for their assistance in combining the PIMD and SIESTA
codes, and A. M. D. Serrano for a critical reading of the manuscript.

\bibliographystyle{apsrev}

\end{document}